\documentclass[pra,twocolumn,showpacs]{revtex4-1}
\usepackage{graphicx,epstopdf}
\usepackage{amsmath}
\usepackage{amssymb}
\usepackage{color}
\usepackage{float}
\usepackage{epsfig}
\usepackage{epstopdf}
\usepackage{framed}
\usepackage{subfigure}
\usepackage{enumerate}

\usepackage[colorlinks=true, citecolor=blue, urlcolor=blue ]{hyperref}

\begin{document}

\title{Transport and localization of light inside a dye-filled microcavity}

\author{Himadri S. Dhar$^1$, Jo{\~a}o D. Rodrigues$^1$, Benjamin T. Walker$^{1,2}$, \\Rupert F. Oulton$^1$, Robert A. Nyman$^1$, and Florian Mintert$^1$}

\affiliation{$^1$Physics Department, Blackett Laboratory, Imperial College London, Prince Consort Road, London SW7 2AZ, UK\\
$^2$Centre for Doctoral Training in Controlled Quantum Dynamics, Imperial College London, Prince Consort Road, London SW7 2AZ, UK}

\begin{abstract}
The driven-dissipative nature of light-matter interaction inside a multimode, dye-filled microcavity makes it an ideal system to study nonequilibrium phenomena, such as transport.
In this work, we investigate how light is efficiently transported inside such a microcavity, mediated by incoherent absorption and emission processes.
In particular, we show that there exist two distinct regimes of transport, viz. conductive and localized, arising from the complex interplay between the thermalizing effect of the dye molecules and the nonequilibrium influence of driving and loss. 
The propagation of light in the conductive regime occurs when several localized cavity modes undergo dynamical phase transitions to a condensed, or lasing, state.
Further, we observe that while such transport is robust for weak disorder in the cavity potential, strong disorder can lead to localization of light even under good thermalizing conditions. 
Importantly, the exhibited transport and localization of light is a manifestation of the nonequilibrium dynamics rather than any coherent interference in the system.
\end{abstract}

\maketitle

\section{Introduction \label{intro}}

In recent years, substantial research effort has been directed to understand the dynamics of both near- and out-of equilibrium systems from the perspective of many-body physics \cite{Polkovnikov2011}, as well as quantum optics \cite{Carusotto2013}. An important 
{phenomenon} related to the study of such dynamics in physical systems is quantum transport. From the flow of electrons in mesoscopic systems \cite{Sohn1997,Datta1997,Sarma2011,Laird2015} and  photons in waveguide quantum electrodynamics (QED) \cite{Shen2007,Hafezi2012,Mahmoodian2018} to energy transfer in molecular systems \cite{Plenio2008,Rebentrost2009} and biological light-harvesting complex in bacteria \cite{Engel2007,Lee2007,Caruso2009}, transport of excitations play a significant role in various areas of natural science. At a fundamental level, transport 
{provides} rich {insight} into the trade-off between several physical processes such as coherent and incoherent interactions, external driving and dissipation.
 On the other hand, the study of transport is also crucial in the design of efficient technologies, ranging from photovoltaic cells \cite{Nelson2003} to quantum networks \cite{Walschaers2013,Mostarda2013}. 

To better understand transport it is necessary to unravel the role of different dynamical processes, and over the past decade, two-dimensional photon gases inside dye-filled microcavities have become useful systems for studying both equilibrium and nonequlibrium physics \cite{Klaers2010b, Marelic2015,Greveling2018,Klaers2010,Schmitt2015,Walker2018,Walker2020}. The photon gas can thermalize via repeated absorption and reemission of photons by the dye molecules, ultimately leading to the formation of a near-equilibrium Bose-Einstein condensate (BEC) \cite{Klaers2010b, Marelic2015,Klaers2010}.
However, the system dynamics is also governed by driven-dissipative processes and the thermalization is ultimately restricted 
by the 
photon losses 
\cite{Marelic2015,Schmitt2015}. This trade-off is notably dependent on the cavity detuning, which can be controlled to investigate the complex interplay between thermalization and loss mechanisms in the dynamics of the photon gas \cite{Kirton2013,Schmitt2014,Kirton2015,Schmitt2016,Keeling2016,Marelic2016,Damm2017,Hesten2018,Walker2019,Walker2019a,Ozturk2019,Vlaho2019,Radonjic2018,Verstraelen2019}. 
Recent experiments using dye-filled microcavities have demonstrated exciting new physics in both these regimes. For instance, exploiting thermalization to create spatially bifurcated coherent quantum states \cite{Kurtscheid2019}, or the study of photon statistics and formation jitter in transient condensation \cite{Walker2020}, and fuzzy phases of photonic condensates \cite{Rodrigues2020}.

In this work, we investigate the transport of light through a lattice of square well potentials inside a multimode, dye-filled microcavity. In particular, we unravel the role of thermalization and nonequilibrium processes in the transport dynamics.
We note that in contrast to transport of quantum excitations or quasiparticles in coherently coupled systems such as electrons in nanostructures or photons in cavity QED, the photon-molecule interaction inside a dye-filled microcavity is mediated by purely incoherent absorption and emission processes.
For fixed driving and losses, the relevant physical parameters here are
the cavity detuning and the structure of the lattice potential. 
The detuning,
$\delta_k = \omega_\mathrm{ZPL}-\omega_k$,
where $\omega_\mathrm{ZPL}$ is the zero-phonon line (ZPL) and $\omega_k$ is the energy of cavity mode $k$, gives the relative strength between rates of absorption, $\mathcal{A}_k$, and emission, $\mathcal{E}_k$,
{following} the Kennard-Stepanov relation \cite{Kennard1918,Kennard1926,Stepanov1957}, $\mathcal{A}_k$ = $\mathcal{E}_k\exp[{-\beta\delta_k}]$. 
The thermalization is then related to the photon absorption in units of the cavity loss rate \cite{Schmitt2015, Keeling2016}. Furthermore, the lattice potential can also be designed to control the spatial overlap of the cavity modes and thus, the effective interaction between the photons and the molecules across the lattice.

The transport problem can then be defined by first considering incoherent excitation
of molecules at the center of the cavity and observing the propagation of light as these excitations gradually spread away from the center, under conditions ranging from well thermalized to strong nonequilibrium regimes.
Second, we also investigate how disorder in the lattice potential affects the cavity modes and therefore the distribution of light in the cavity plane. 
While transport of light occurs for very weak disorder, localization is expected when the disorder is strong.
We note that any phenomena of transport or localization of light observed here are intrinsic characteristics of the transition from near- to out-of-equilibrium dynamics, rather than any weak quantum interference \cite{Datta1997} or Anderson localization \cite{Anderson1958,Abrahams1979}.

The paper is arranged as follows. We present the rate equations for the driven-dissipative system in Sec.~\ref{model}. Based on these equations, in Sec.~\ref{trans} we show the transport of the photon density. In Sec.~\ref{phase}, we derive an effective model to study transport and show the emergence of distinct regimes of transport. In Sec.~\ref{local}, we look at the effect of disorder in the lattice on the properties of transport. We end with a discussion in Sec.~\ref{disc}.

\section{Driven-dissipative system \label{model}}

The light-matter interaction inside the dye-filled microcavity and the corresponding driven-dissipative dynamics can be studied using 
{a nonequilibrium model~\cite{Kirton2013}, which results in the Markovian master equation~\cite{Keeling2016}}
\begin{eqnarray}
\frac{d\rho}{dt} &=& -i[H_0,\rho] - \sum_{k,i} \left\{\kappa \mathcal{L}[\hat{a}_k] + \Gamma_\uparrow(\mathbf{x_i}) \mathcal{L}[\hat{\sigma}^+_i]  +  \Gamma_\downarrow \mathcal{L}[\hat{\sigma}^-_i] \right.\nonumber\\
&+& \left.\lvert \Psi_k(\mathbf{x_i}) \rvert^2(\mathcal{A}_k \mathcal{L}[\hat{a}^\dag_k\hat{\sigma}^-_i] + \mathcal{E}_k \mathcal{L}[\hat{a}_k\hat{\sigma}^+_i])
\right\} \rho.
\label{noneq_model}
\end{eqnarray}
Here, $\hat{a}_k$ ($\hat{a}^\dag_k$) is the photon annihilation (creation) operator for the $k^{th}$ cavity mode {and $\hat{\sigma}^{\pm}_i$} is the Pauli operator corresponding to the $i^{th}$ molecule. 
The rate of photon loss and {molecular decay} into noncavity modes are $\kappa$ and $\Gamma_\downarrow$, respectively.
Further, $H_0 = -\delta_k \hat{a}^\dag_k\hat{a}_k$ is the Hamiltonian in the limit of weak-coupling \cite{Keeling2016} and the Lindblad operator is $\mathcal{L}[\hat{X}]\rho$ = $\frac{1}{2}\left\{\hat{X}^\dag\hat{X},\rho \right\} - \hat{X}\rho\hat{X}^\dag$. Also, $\Gamma_\uparrow(\mathbf{x_i})$ is the pump rate to excite molecules at position $\mathbf{x_i}$.

\begin{figure}[t]
\epsfig{figure = 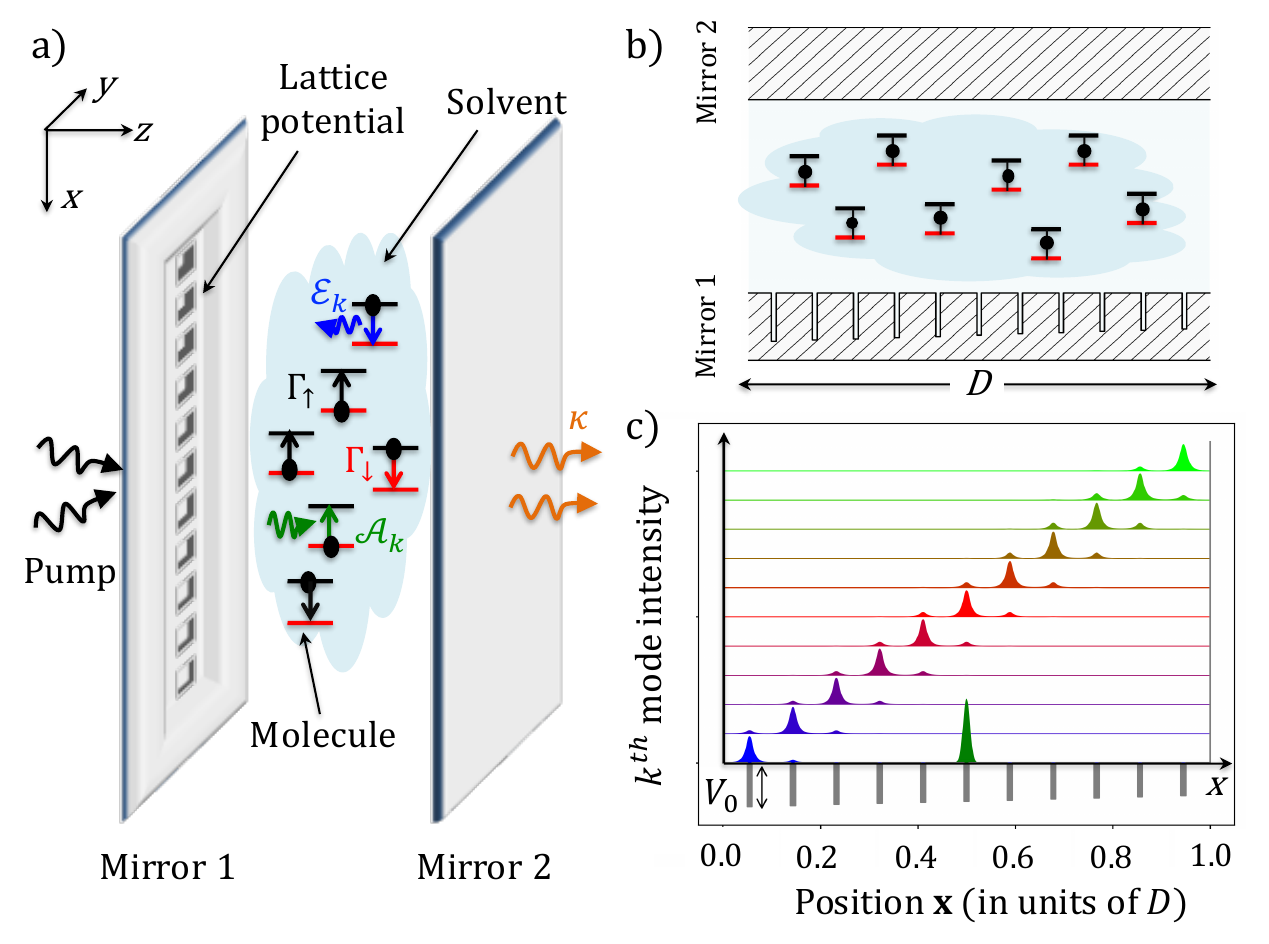, width=0.49\textwidth,angle=-0}
\caption{(Color online.) {Light-matter interaction in a dye-filled microcavity.} a) A schematic showing a few two-level dye molecules in the solvent, placed between the mirrors of a microcavity, along with a depiction of the different incoherent processes taking place. b) A sketch showing how the potential lattice of width $D$ can be fabricated on the transverse space of the mirror.
c) The figure shows the normalized mode intensities in the transverse space of the cavity, {for a square well lattice with $s=11$ sites.
The potential is biased, which results in a set of low-energy cavity modes that are localized around each site. 
The $k$ mode intensities are shown along the vertical axis, whereas, the horizontal x-axis is divided into $S$ spatial bins.
The dye molecules at the central lattice site are excited using an external pump (shown in green).}
}
 \label{fig_modes}
\end{figure}

The mode function of the $k^\mathrm{th}$  cavity mode, $\Psi_k(\mathbf{x_i})$, in Eq.~(\ref{noneq_model}) can be obtained by solving the Schr{\" o}dinger equation for the potential in the transverse space of the microcavity. 
For instance, for the harmonic oscillator potential used in several experiments on photon BEC \cite{Klaers2010b, Marelic2015,Klaers2010}, these functions are Hermite polynomials. 
Here, we consider a  transverse potential landscape consisting of a one-dimensional (1D) lattice of square well potentials.  This can be realized by fabricating quasi 1D potentials on the planar substrate of dielectric mirrors using focussed ion beam milling \cite{Flatten2016,Trichet2015,Walker2020b} (see Fig.~\ref{fig_modes}). We consider a trapping potential of total width $D$ on the mirror, which contains $s$ square wells, each of width, $\Delta w$, and separated by $\Delta d$. The potential depth of each well is $V_l$ = $V_0 - \Delta V_l$, where $l = 0,1,\dots,s-1$ and $\Delta V_l$ is the variation in potential arising from any uniform bias or disorder in the potential. 
The intensity of the $k^\mathrm{th}$ mode function at position $\mathbf{x}$ is given by $g_k(\mathbf{x}) = \lvert\Psi_k(\mathbf{x})\rvert^2$, {as shown in Fig.~1(c)}, which is dependent on the dimensions of the potential lattice, as defined by $V_l$, $\Delta w$ and $\Delta d$.

\begin{figure*}[t]
\epsfig{figure = 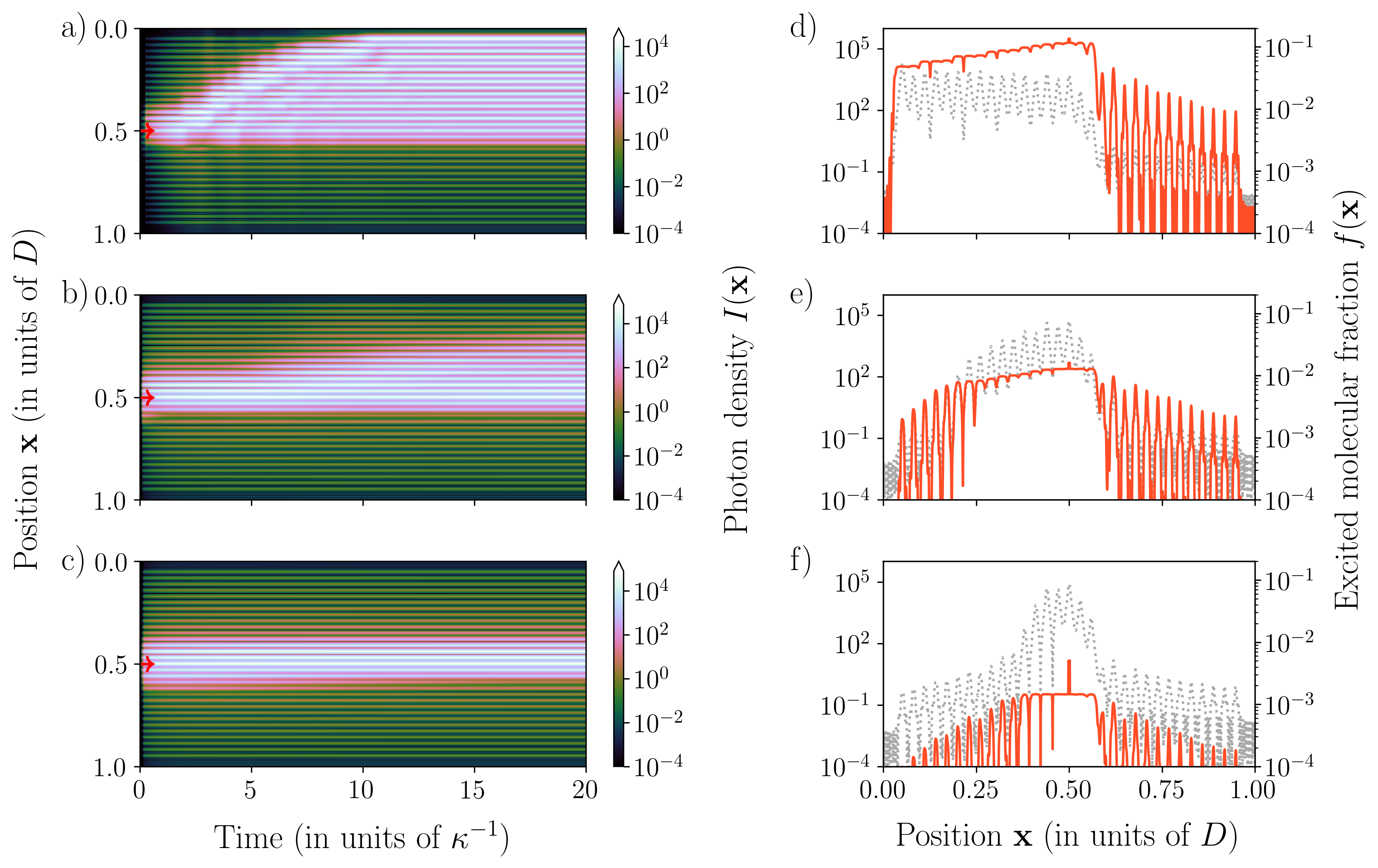, width=0.7\textwidth,angle=-0}
\caption{(Color online.) Transport of light in a biased lattice potential. The figures on the left shows the flow of photon density, $I(\mathbf{x})$, across the lattice for different values 
of the cavity detuning, $\delta_0$ (in units of $10^{-2}\omega_\mathrm{ZPL}$): a) 3.4, b) 6.0, and c) 8.6.
The lattice potential has wells with a fixed width, $\Delta w/D = 0.4\times10^{-2}$, and is biased to the left as shown in Fig.~\ref{fig_modes}.  The red arrow shows the center of the cavity where the molecules are incoherently pumped. 
The figures on the right show the spatial distribution of excited molecules, $f(\mathbf{x}_j) = m_j/M$ (red-solid), and the photon density, $I(\mathbf{x})$ (black-dotted), at steady state, for the same set of values of the cavity detuning, $\delta_0$, in the left plots. {Here, $D = 100~\mu$m, $\kappa = 50$ THz, and $p = 40~\Gamma_\downarrow$.}
}
 \label{fig_trans}
\end{figure*}

To numerically solve Eq.~(\ref{noneq_model}), we first consider the semiclassical approximation that is valid for a large number of molecules, where all photon-molecule coupling can be factorized, i.e., $\langle\hat{a}_k\sigma^+_i\rangle = \langle\hat{a}_k\rangle\langle\sigma^+_i\rangle$. Secondly, the effect of mode coherence is assumed to be small compared to the population for the largely incoherent long-time and steady-state dynamics. Therefore, a closed set of rate equations
for the photon numbers, $n_k = \langle\hat{a}^\dag_k\hat{a}_k\rangle$, can be obtained. 
For a spatially resolved distribution of molecules \cite{Keeling2016,Walker2018,Walker2019a}, these are given by
%
%
\begin{eqnarray}
\label{eq:ndot}
\frac{dn_k}{dt} &=& - \kappa n_k + \sum_{j=1}^S g_{k,j} \left[\mathcal{E}_k m_j \left( n_k + 1 \right)\right. \nonumber\\
&-&\left.\mathcal{A}_k  \left(M - m_j \right) n_k \right].
\end{eqnarray}
In the above equation, the 1D transverse space of the cavity is divided into $S$ spatial bins, such that each bin contains $M$  = $N/S$ dye molecules, where $N$ is the total number of molecules in the cavity. {Moreover, $m_j$ = $\sum_{i=1}^{M} \sigma_i^+(\mathbf{x}_j)\sigma_i^-(\mathbf{x}_j)$, is the number of excited molecules 
in the $j^\mathrm{th}$ spatial bin.} 
{The amplitude $g_{k,j} = \lvert \Psi_k(\mathbf{x_j}) \rvert^2$ denotes the overlap} between the mode $k$ and the molecules in the spatial bin $j$, located at $\mathbf{x_j}$.
%
Now, the set of rate equations for the molecular excitation, $m_j$, in each spatial bin is given by,
\begin{eqnarray}
\frac{d{m}_j}{dt} &=& - \{\Gamma_\downarrow + \sum_k \mathcal{E}_k g_{k,j}  \left(n_k + 1 \right)\}m_j \nonumber\\
&+& \{\Gamma_\uparrow(\mathbf{x_j}) + \sum_k \mathcal{A}_k g_{k,j}  n_k\} (M - m_j),\label{eq:fdot}\\
&=& -\Gamma^\mathrm{tot}_\downarrow m_j + \Gamma^\mathrm{tot}_\uparrow (M - m_j).
\end{eqnarray}
From the set of semiclassical Eqs.~(\ref{eq:ndot}-\ref{eq:fdot}), the
quantities of interest are the spectrum of photon numbers $n_k$ and the photon density in the transverse space of the cavity, given by, 
${I}(\mathbf{x}) = \sum_k \lvert\Psi_k(\mathbf{x})\rvert^2 n_k$ \cite{Keeling2016}. The transport and localization of light in the lattice can then be measured in terms of the 
propagation of the photon density, ${I}(\mathbf{x})$.

The transport of light inside the dye-filled microcavity is primarily dependent on two key processes. First, the balance between thermalization and nonequilibrium processes, which depends on the cavity detuning, $\delta_0$, the photon loss rate, $\kappa$ and the pump rate, $\Gamma_\uparrow(\mathrm{x})$. Second, the structure and dimension of the lattice potential in the transverse space of the cavity, which determines the spread or spatial overlap of the mode intensities, i.e., how much do modes interact with shared molecular populations at distant lattice sites. To account for the effect of these factors, we investigate two distinct processes: 

1) \emph{Thermalization-assisted transport}: For a linear potential bias in the lattice (such that the leftmost site is energetically the most favorable) and by adjusting the well-width, $\Delta w$, the spread of the cavity modes are limited to few neighboring sites (see Fig.~\ref{fig_modes}). The transport in this case is governed purely by dynamical processes.

2) \emph{Disorder induced localization}: For an ordered equipotential lattice, the cavity modes are delocalized thus allowing for a direct transport of the photons. However, a small disorder can break this delocalization and the light propagation then depends on the relative strengths of thermalization and disorder.

\begin{figure}[t]
\epsfig{figure = 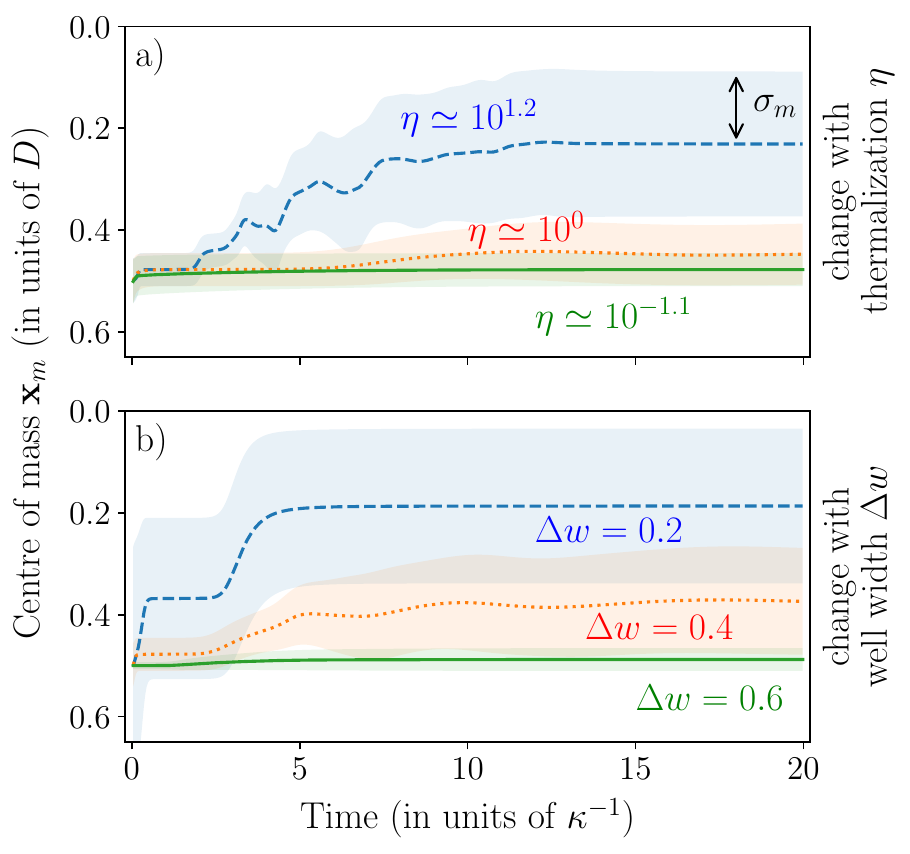, width=0.45\textwidth,angle=-0}
\caption{(Color online.) The transport of the center of mass. The figures shows the evolution of the center of mass, $\mathbf{x}_m$, of the photon density, $I(\mathbf{x})$. The shaded region is the standard deviation or width, $\sigma_m$, of the photon density, $I(\mathbf{x})$ (black arrow). The plots are a) for fixed well-width, $\Delta w$ = $0.4$ (in units of $D\times10^{-2}$) and varying thermalization coefficient, $\eta$, and b) for fixed $\eta = 10^{1/2}$, and increasing well-width, $\Delta w$. 
{The remaining parameters are the same as Fig.~\ref{fig_trans}.}}
\label{fig_com}
\end{figure}

\section{Thermalization-assisted transport \label{trans}}

In a dye-filled microcavity, the thermalization of the photon gas {at room-temperature}  is a result of fast and repetitive absorption and re-emission of photons by the molecules.
A key 
{criterion} here is that the dye molecules absorb and re-emit light in the cavity at a rate faster than the loss of photon from the cavity. 
The thermalization can be quantified in terms of a dimensionless coefficient, $\eta = \mathcal{A}_0M/\kappa$ \cite{Keeling2016,Hesten2018}, where  $\mathcal{A}_0$ 
is the absorption rate of the lowest-energy cavity mode.
For gain media typically used in photon BEC experiments, 
the absorption rate in the operational regime decreases with increased cavity detuning, $\delta_0$ = $\omega_\mathrm{ZPL} - \omega_{0}$, where {$\omega_\mathrm{ZPL}$ is the zero-phonon line and $\omega_{0}$ is the cavity cutoff frequency.}
Therefore, thermalization in the system weakens ($\eta$ decreases) with increase in either the detuning, $\delta_0$, or the cavity loss, $\kappa$.


{In Figure~\ref{fig_trans}, transport of light is studied in a transverse square well lattice with a linear potential bias, $V_l$ = $V_0 - l \Delta V$, where $l = 0,1,\dots,30$ are the lattice sites. The potential, $V_0 \approx 10^{4}$ (in units of $\hbar c_0^2/\omega_0 D$), with $\Delta V/V_0 = 4 \times 10^{-3}$.  Here, $c_0$ is the speed of light inside the cavity and $D$ is the lattice width.}
The system is excited by incoherently driving or pumping the molecules close to the central lattice site {at $\mathbf{x}_{c}$, i.e., $p = \Gamma_\uparrow(\mathbf{x_c})$. The molecular decay rate is fixed at, $\Gamma_\downarrow =  \kappa/8$, where $\kappa$ is the cavity loss rate. The absorption ($\mathcal{A}_k$) and emission ($\mathcal{E}_k$) rates are derived from experimental spectra for Rhodamine 6G dye-molecules \cite{rhoda6g}, which has a zero-phonon line, $\omega_\mathrm{ZPL} = 550$~THz.}
The figure shows the flow of the photon density, $I(\mathbf{x})$, as obtained from the rate equations in Eqs.~(\ref{eq:ndot}-\ref{eq:fdot}), for different thermalization conditions. For a biased potential and a fixed well-width, $\Delta w = 4\times10^{-3}D$, {the intensity of mode $k$ is mainly localized around the lattice site $l=k$, as shown in Fig.~\ref{fig_modes}(c). This implies that $g_{k,j} \approx 0$ for all spatial bins $\mathbf{x}_j$ that do not belong to lattice sites,  $l = k,k\pm1$. In other words, the cavity modes at a lattice site do not significantly couple with molecules located beyond its nearest-neighbor (NN) lattice sites.}
After pumping, the emitted photon density is initially centered around the pump spot, 
and with increasing time,
the photon density, $I(\mathbf{x})$, is shown to propagate towards the left-edge of the lattice, which is energetically more favorable in the biased lattice potential. It is observed that transport is efficient for low cavity detuning, $\delta_0$, where the system is well thermalized. 

To understand the spread of light, two relevant figures of merit are the center of mass, $\mathbf{x}_m$, and the standard deviation or width, $\sigma_m$, of the photon density, $I(\mathbf{x})$. These can be defined as 
\begin{eqnarray}
\mathbf{x}_m &=& \frac{1}{\mathcal{I}}\sum_{i=1}^N I(\mathbf{x}_i)~  \mathbf{x}_i, ~~~\mathrm{and} \\
\sigma_m^2 &=& \frac{1}{\mathcal{I}} \sum_{i=1}^N I(\mathbf{x}_i) ~(\mathbf{x}_i - \mathbf{x}_m)^2,
\end{eqnarray}
where $\mathcal{I}$ = $\sum_i I(\mathbf{x}_i)$, is the intensity of light in the cavity. 
In Fig.~\ref{fig_com}(a), we observe how these quantities evolve for different thermalization coefficients, $\eta$,  corresponding to the cavity detunings, $\delta_0$, used in Fig.~\ref{fig_trans}.
For  $\eta \gg 1$, the center of mass, $\mathbf{x}_m$, moves towards the left of the pump spot, and has a large width, $\sigma_m$. This implies that  sites at the edge of the cavity are occupied, which is consistent with the transport of $I(\mathbf{x})$ in Fig.~\ref{fig_trans}. 
The evolution of $\mathbf{x}_m$ is also dependent on the structure of the lattice, as shown in  Fig.~\ref{fig_com}(b). For a fixed $\eta$, the flow of $\mathbf{x}_m$ is more prominent for 
smaller $\Delta w$, where the cavity modes overlap beyond NN lattice sites, thus allowing for more efficient transport. On the other hand, for modes that have no or minimal overlap with any other lattice site, $\mathbf{x}_m$ is now restricted to the center of the cavity, $\mathbf{x}_{c}$, with a small width, $\sigma_m$.

The transport of the photon density can be explained in terms of the macroscopic occupation of the modes at each site, starting from the pump spot to the edge of the lattice. 
As the dye molecules are incoherently pumped, the cavity mode closest to the pump spot at the center undergoes a dynamical phase transition from a disordered, low photon-number phase to a macroscopically occupied, condensed or lasing phase \cite{Walker2020}. Below the condensation threshold, the photon intensity, $I(\mathbf{x})$, in the cavity is low and follows a Gaussian distribution with a peak at the center, and intensity at distant site $m$ tailing off as $e^{-|\omega_m - \omega_c|}$, where $\omega_c$ is the frequency of the pumped mode.
{Figures~\ref{fig_trans}(a)-(c) shows that the intensity at a condensed mode is typically $10^4$ times greater than the value below threshold.}
Above threshold, incoherent interactions allow excitations to
spread across the lattice and in a well thermalized regime, with increasing time successive modes away from the pump spot undergo phase transitions, till the lowest-energy mode at the left-most lattice site is macroscopically occupied. For poor thermalization, the condensation of modes does not extend beyond a certain lattice site. 
The effect of thermalization in the system is also evident in Figs.~\ref{fig_trans}(d)-(e), which shows the spatial spread of the  molecular excitation fraction in the lattice for different cavity detunings. For condensed modes, the large photon population clamps the excitation fraction to a fixed value \cite{Keeling2016,Hesten2018}.
For low detuning, $\delta_0$, the clamped excitation is spread wide from the center to the left-edge of the cavity where the modes have condensed, in contrast to high detuning, where the condensation and clamping is restricted to the pump spot.

\begin{figure*}[t]
\epsfig{figure = 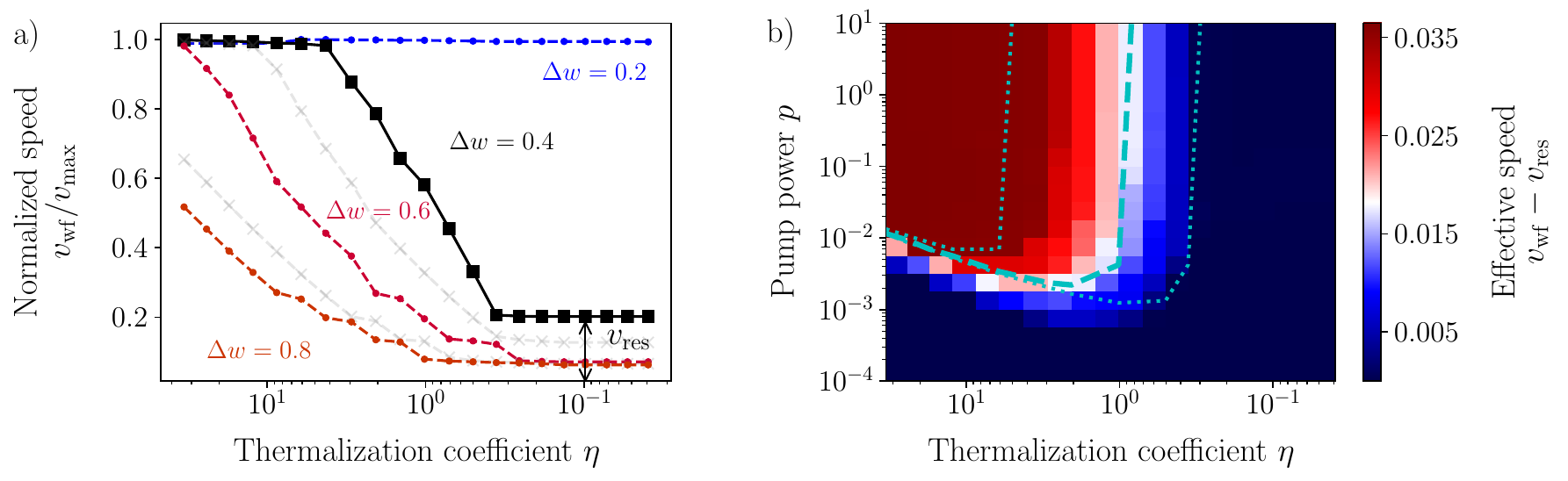, width=0.8\textwidth,angle=-0}
\caption{(Color online.) The speed of transport of the light intensity. a) The plot shows the normalized average speed of the light intensity wavefront, $v_\mathrm{wf}/v_\mathrm{max}$, for different values of thermalization coefficient, $\eta$, and well-width, $\Delta w$ (in units of $D\times10^{-2}$).  A couple of intermediate values of $\Delta w$ are shown in light-gray. 
Here, $v_\mathrm{res}$ (black arrow) is the residual velocity at low $\eta$ arising from transport simply due to the spatial overlap of the cavity modes.
b) The phase diagram for thermalization-assisted transport of light shows the
effective speed, $v_\mathrm{wf} - v_\mathrm{res}$ for different $\eta$, and pump rates, $p$  (in units of $ 4\times10^{3}\Gamma_\downarrow$). The well-width is fixed at $\Delta w = 0.4$ (black-solid-square in the left figure). The boundary predicted by the effective model is shown by a blue-bold-dashed line. The boundaries for lesser or greater overlap of the cavity mode are shown with blue-dotted lines.
}
 \label{fig_speed}
\end{figure*}

\section{Regimes of light transport \label{phase}}

%
A natural way to classify 
thermalization-assisted transport of light, 
is to develop an effective model that clearly explains the dynamical phase transitions in the system.
%
To this end, we now consider a closed, simplified form of the nonequilibrium model. 
As a start, the spatial bins in the transverse space is taken to be equal to the number of sites or wells in the lattice potential, i.e., $S = s$.
{The cavity mode $k$ is then localized at $j = k, k\pm1$, such that it only overlaps with NN sites: $g_{k,j} \neq 0~\forall~j = k, k\pm1$.}
Excitation to all other regions are mediated via the dynamical processes, i.e., the absorption and emission process. Hence, molecules in 
non-pumped bins are excited only after the cavity mode at {the} central lattice site undergoes a condensation phase transition and is macroscopically occupied. Further, assuming a potential bias to the left of the lattice, mode $k$ is only occupied after the mode $k+1$ has condensed.
As such, the rate equation for the molecular population in the $k^\mathrm{th}$ spatial bin can be written as ($\Gamma_\uparrow(\mathbf{x_k})$ = $\Gamma_\downarrow = 0$)~:
\begin{eqnarray}
\frac{d{m}_k}{dt} &=& \mathcal{A}_{k+1} g_{k+1,k} n_{k+1} (N_k - m_k) - \mathcal{E}_{k+1} g_{k+1,k} n_{k+1} m_k\nonumber\\
&=& \mathcal{G}_{k+1} (N_k - \nu_{k+1} m_k), 
\label{simp_mod}
\end{eqnarray}
where, $\mathcal{G}_{k+1}$ = $\mathcal{A}_{k+1} g_{k+1,k} n_{k+1}$ and  $\nu_{k} = 1 + e^{\beta \delta_{k}}$.
{Here, $n_{k+1}$ is the photon number of the condensed mode $k+1$. For simplicity, we set $n_{k} =  n_s$ for all $k$ modes that have condensed, where $n_s$ is an estimate of the macroscopic occupation.} All other quantities are the same as defined earlier.
Solving Eq.~(\ref{simp_mod}), the time taken to reach the excitation $m_k$ is given by
\begin{eqnarray}
t_k = \frac{1}{\mathcal{G}_{k+1}\nu_{k+1}} \log_\mathrm{e} \left( \frac{N_k}{N_k-\nu_{k+1}m_k} \right).
\end{eqnarray}
Now, for  mode $k$ to undergo a phase transition to a condensed state, the molecular population should be gain-clamped \cite{Hesten2018}, which for $n_k \gg 0$, is given by
\begin{eqnarray}
{m_k} &=& \dfrac{M \Gamma^\mathrm{tot}_\uparrow}{\Gamma^\mathrm{tot}_\uparrow+\Gamma^\mathrm{tot}_\downarrow} =  \frac{N_k \mathcal{A}_k}{ \mathcal{A}_k+ \mathcal{E}_k} = \frac{N_k}{\nu_{k}},~
\mathrm{which~gives~} \\
t_k &=& \frac{1}{\mathcal{G}_{k+1}\nu_{k+1}} \log_\mathrm{e} \left( \frac{1}{1-\nu_{k+1}/\nu_{k}} \right). \label{time_model}
\end{eqnarray}
%
%
Therefore, $t_k$ is the time taken for the cavity mode $k$ to condense and is the approximate time taken for light to transport from lattice site $k+1$ to $k$.
For cases where transport breaks down, $t_k$, is typically large and comparable to the losses, $\kappa$ and $\Gamma_\downarrow$ in the system. 
The relevant point is that $t_k$ is inversely proportional to $\mathcal{G}_{k+1}$, which directly depends on the absorption, $\mathcal{A}_{k+1}$ and the spatial overlap, $g_{k,k+1}$, thus giving us the expected dependence on thermalization and structure of the lattice potential.
In other words, the transport time is small for higher absorption rate, $\mathcal{A}_{k+1}$, which is the very root of
thermalization assisted transport as shown in Figs.~\ref{fig_trans}-\ref{fig_com}. Similarly, $t_k$  is also small for higher mode overlap, $g_{k+1,k}$, which depends on the structure of lattice potential, and leads to propagation of light as observed in Fig.~\ref{fig_com}. Therefore, the model provides a reliable and efficient mechanism to explain the transport of light inside a dye-filled microcavity based on a series of dynamical phase transitions in neighboring cavity modes.

\begin{figure*}[t]
\epsfig{figure = 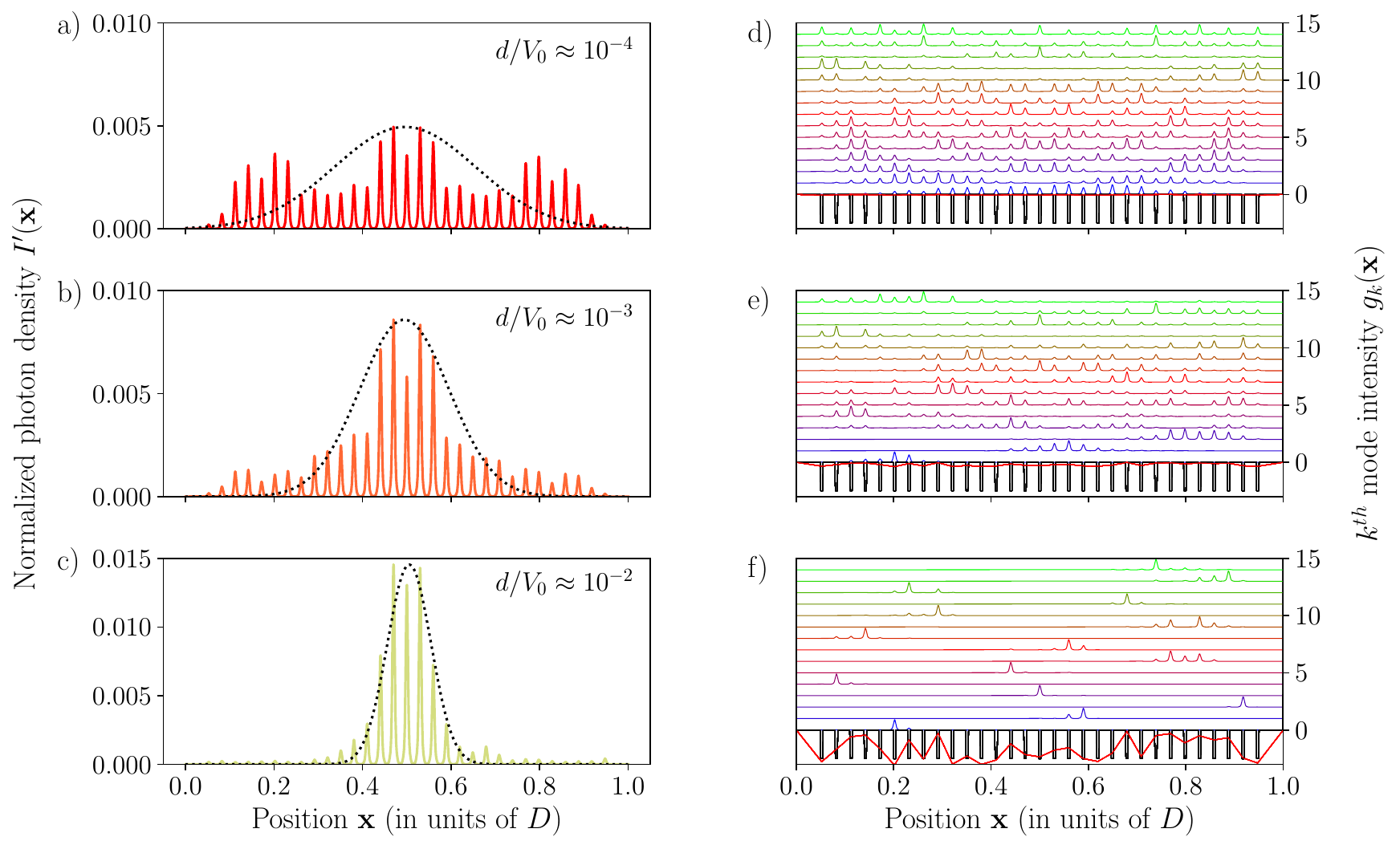, width=0.75\textwidth,angle=-0}
\caption{(Color online.) Light in a disordered lattice potential. The figures on the left shows the spatial distribution of the normalized photon density, $I'(\mathbf{x}) = I(\mathbf{x})/\mathcal{I}$,
at steady state, for different degrees of disorder, $d/V_0$: a)  $10^{-4}$, b) $10^{-3}$, and c) $10^{-2}$. The dotted lines represent a Gaussian envelope with the same peak and standard deviation as that of the normalized photon density. Here, the thermalization coefficient, $\eta$, and well-width, $\Delta w$, are fixed for all plots. d)-f) The figures on the right shows the low-energy cavity mode intensities, $g_k(\mathbf{x}) = \lvert\Psi_k(\mathbf{x})\rvert^2$, corresponding to the above disordered lattice potential. The square well potential at each lattice site (black) and a scaled visualization of the  disorder in the lattice energy (red) are also shown. 
}
 \label{fig_loc}
\end{figure*}

To compare the model with the explicit numerical calculations of the driven-dissipative rate equations in Eqs.~(\ref{eq:ndot}-\ref{eq:fdot}), the temporal evolution of the photon density wave inside the cavity needs to be captured.
A relevant quantity is the front of the propagating photon density wave, $\mathbf{x}_\mathrm{wf}(t)$ at time $t$, which is defined as the position in the transverse space where  99\% of the total light intensity \cite{comment1} in the cavity is to its right, i.e.,
\[
\mathbf{x}_\mathrm{wf}(t) : \sum_{\mathbf{x}_i>\mathbf{x}_\mathrm{wf}} I(\mathbf{x}_i)(t) \approx 0.99 ~\mathcal{I}(t)
\]
At time $t = T$, one can define an average speed with which this intensity wavefront, $\mathbf{x}_\mathrm{wf}(t)$, moves away from the center of the cavity, $\mathbf{x}_{c}$, such that
\begin{equation}
v_\mathrm{wf} = \left.\frac{\Delta \mathbf{x}_\mathrm{wf}(t)}{\Delta t} \right\vert_{t=0}^T, 
\label{speed_sim}
\end{equation}
where, $\Delta \mathbf{x}_\mathrm{wf}(t) = \mathbf{x}_\mathrm{wf}(t) - \mathbf{x}_{c}$.
The average transport speed of photon density can also be obtained from our effective model using Eq.~(\ref{time_model}):
\begin{eqnarray}
v_\textrm{mod} &=& \left. \frac{\Delta x(t)}{t} \right|_{t = T} = \frac{L(T)\Delta d}{T}, 
\label{speed_mod}
\end{eqnarray}
where, $t = \sum t_k + t'$. Here, $L(T)$ is the number of local modes that have condensed at $t = T$ and $\Delta d$ is the space between each site. Here, $t'$ is the time taken to gain-clamp the molecules at the pumped site, $k=k_c$, which can be obtained by solving:
\begin{eqnarray}
&\dot{m}_{k_c}  = -(\Gamma_\downarrow + \mathcal{E}_{k_c}) m_{k_c} + \Gamma_\uparrow (M -m_{k_c});~~n_{k_c} \approx 0,&\nonumber\\
&t' = \dfrac{1}{\Gamma_\uparrow}\log_\mathrm{e}\left( \dfrac{1}{1-\gamma/\nu_{k_c}} \right),~ \mathrm{where}~~\gamma = \dfrac{p}{p+\Gamma_\downarrow+\mathcal{E}_{k_c}}.&\nonumber
\end{eqnarray}

Figure~\ref{fig_speed}, shows the average speed, $v_\mathrm{wf}$, with which the intensity wavefront propagates inside the cavity as a function of the thermalization coefficient, $\eta$, the pump power, $p$, and the potential well-width at each lattice site, $\Delta w$. 
The normalized speed, $v_\mathrm{wf}/v_\mathrm{max}$, where $v_\mathrm{max}$ is the maximum speed at time $T$, is shown in Fig.~\ref{fig_speed}(a). This is	equal to unity when the  mode overlap is large for small $\Delta w$. On the other hand, the speed is less for relatively large values of $\Delta w$.
%
Importantly, for intermediate values of $\delta w$, where the mode overlap is mostly limited to NN sites, the plot highlights two distinct regimes of transport, namely the 
{conductive} regime, where the intensity wavefront follow the potential bias and propagates  to the leftmost site of the lattice, and the {localized} regime, where the light is concentrated around the pump spot and the wavefront speed, $v_\mathrm{wf}$, is equal to the residual speed, $v_\mathrm{res}$, arising simply from the overlap of the cavity modes.

This is further demonstrated in Fig.~\ref{fig_speed}(b), which shows the phase diagram for thermalization-assisted speed, $v_\mathrm{wf}-v_\mathrm{res}$. Here, the parameter space spanned by the thermalization, $\eta$ and pump power, $p$ is demarcated into the conductive and localized transport regimes. The figure also shows that the boundary separating the regimes is quite accurately predicted by the average speed calculated in Eq.~(\ref{speed_mod}), using the effective model, for the parameters, {$g_{k+1,k} = 0.1$, $n_{s} \approx 5\times10^4$, and $\Delta d = 0.03~D$.} 
Additionally, the shift in the boundary as the mode overlap, $g_{k+1,k}$, changes is also shown in the phase-diagram. Within the realms of the effective model, the phase boundary moves to the right for $g_{k+1,k} = 0.33$, to include light in less thermalized regions, in contrast to the case, $g_{k+1,k} = 0.01$, where higher thermalization is necessary for transport.

\section{Disorder induced Localization\label{local}}

While {the} thermalization-assisted transport model derived in Sec.~\ref{phase} explains the propagation of light via 
the successive condensation of neighboring lattice sites,
this may no longer be qualitatively accurate in the presence of disorder. Apart from its importance in the study of nonequilibrium processes, disorder in the lattice energy is a crucial factor in any physical implementation of the system, as
the transition from an ordered to a disordered potential landscape will substantially affect the spatial distribution of the cavity modes, $g_{k,j}$, and therefore the photon density, $I(\mathbf{x})$.
To study this transition, we start with an unbiased and ordered lattice on the transverse space, with potential $V_0$ at each site. For such a symmetric lattice potential, the cavity modes are uniformly spread across all sites (with some boundary effect) and therefore,
{have} finite overlap with all molecular spatial bins, i.e., $g_{k,j} > 0,~\forall~j$.
All emission to the cavity modes therefore results in the photon density, $I(\mathbf{x})$, being near-uniformly distributed in the transverse space. When the dye molecules at the center of the cavity are incoherently pumped,
and in the absence of any potential bias, the center of mass, $\mathbf{x}_{m}$, of the photon density is expected to remain close to the pump spot due to light propagating along both directions. 
The key quantity here is the spatial spread of light in the lattice as quantified by the standard deviation or width, $\sigma_m$, of the photon density, $I(\mathbf{x})$.
The disorder in the potential is implemented by randomly changing the energy at each lattice site, such that 
$V_l = V_0 - \Delta V_l$, where
$\Delta V_l$ = $x_l d$ and $x_l \in [-1/2,1/2]$ is a random number. Here, $d$ is the degree of disorder ranging from 0 to $10^{-1}V_0$.
For finite $d/V_0$, the cavity modes begin to localize around a few lattice sites.

\begin{figure}[t]
\epsfig{figure = 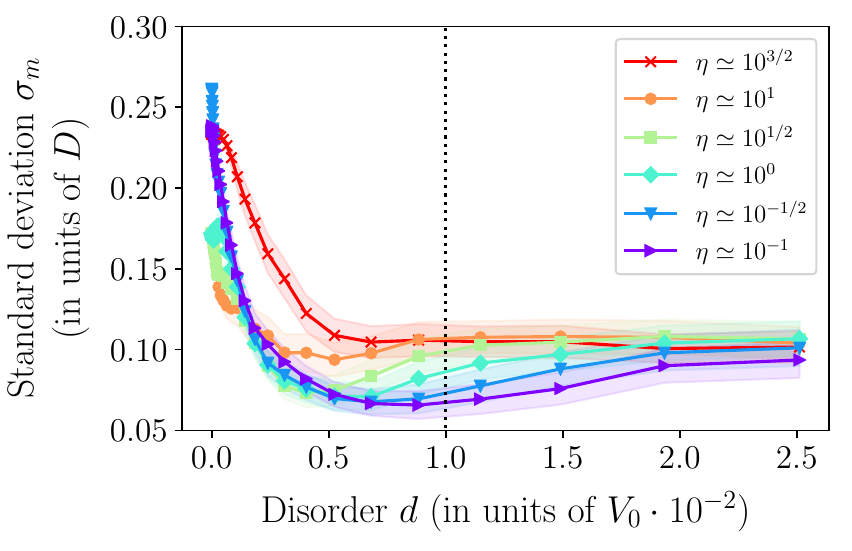, width=0.45\textwidth,angle=-0}
\caption{(Color online.)  Localization of light in a disordered lattice potential. In the figure, we show the standard deviation or width, $\sigma_m$, of the steady-state photon density, $I(\mathbf{x})$, for different values of the degree of disorder $d$ and thermalization coefficient, $\eta$. Here, $\sigma_m$ is estimated by averaging over 200 realizations of the disorder. The shaded region highlights the 99\% confidence interval in calculating the width, $\sigma_m$.
}
 \label{fig_locwid}
\end{figure}

Figure~\ref{fig_loc}, shows the spread of the normalized photon density, $I'(\mathbf{x}) = I(\mathbf{x})/\mathcal{I}$, at steady state, for a set of disordered lattice potentials. Here, the steady state of light inside the cavity is averaged over $200$ sets of random lattice potentials for each value of degree of disorder, $d$.
The thermalization is fixed at $\eta \approx 1.0$ and the well-width at each site is, $\Delta w = 5\times10^{-3}D$. The disorder in the potential increases as $d$ changes from $10^{-4}V_0$ to $10^{-2}V_0$ in Fig.~\ref{fig_loc}(a)-(c). 
For very low or close to negligible disorder in the system the normalized photon density, $I'(\mathbf{x})$, is well distributed across the lattice, as the low-lying cavity modes are well spread across the lattice.  For higher disorder,  $I'(\mathbf{x})$ is increasingly localized along the center of the cavity, where the molecules were initially pumped. 
The distribution of photon density is connected to the spatial spread of the low-energy cavity modes in the disordered system, as shown in Fig.~\ref{fig_loc}(d)-(f). As $d/V_0$ increases, the low-cavity modes tend to be more concentrated around a few sites in the lattice.

To quantitatively study the distribution of light in disordered lattice potentials, we look at the standard deviation or width, $\sigma_m$, of the photon density, $I(\mathbf{x})$. For instance, in Fig.~\ref{fig_loc}(a)-(c), the distribution of light is also highlighted by the Gaussian envelope with the same peak and width, $\sigma_m$, as for the normalized photon density.
In Fig.~\ref{fig_locwid}, we analyze the width, $\sigma_m$, of the steady-state photon density, $I(\mathbf{x})$, for different values of thermalization, $\eta$ and disorder, $d$.
The figure shows that the width, $\sigma_m$, is large for very low disorder, $d \approx 0$, which implies the absence of any localization effect. Here the thermalization in the system appears to not play any significant role as the low-energy cavity modes are delocalized. However, as disorder increases, the width, $\sigma_m$, reduces and the photon density begins to localize around the pump spot.

The random lattice potential provides some interesting insights into the interplay between thermalization, nonequilibrium processes and disorder in the system. For very low disorder degree, $d$, the low-energy cavity modes are spread across several lattice sites. The light propagates simply through these delocalized modes unhindered by any dynamical process. As the disorder increases, the overlap of the cavity mode with the lattice is restricted in space, and the absorption and emission of light by the dye-molecules and other nonequilibrium processes assume greater importance. As shown in Fig.~\ref{fig_locwid}, for  
$0 < d/V_0 < 1.0 \times 10^{-2}$, the width, $\sigma_m$, is higher for better thermalization regimes, $\eta > 1$, which implies the system exhibits some form of thermalization-assisted transport, i.e., the propagation of light in this regime should be mediated by transitions to the condensed or lasing state as discussed in Sec.~\ref{phase}.

Surprisingly, the transport does not monotonically decrease with increasing disorder. For all thermalization values there exists a parameter range where the width,  $\sigma_m$, anomalously increases with increasing disorder, $d$. This is due to the fact that as disorder is increased in this regime, the randomness may lead to cavity modes at distant sites having a more favorable absorption rate. These sites are more likely to be macroscopically occupied and allows on average for greater spread of light, provided some delocalization of cavity modes persist in this regime.
As $d$ increases further, the cavity modes have reduced spatial overlap with neighboring sites and the probability of finding desirable transition condition is lost. Therefore,  the transport is restricted regardless of the near or out-of equilibrium conditions. For $d/V_0 > 2.0 \times10^{-2}$, as shown in Fig.~\ref{fig_locwid}, the light is localized close to the pump spot for all values of the thermalization coefficient, $\eta$.

\section{Discussion \label{disc}}

The light-matter interaction in a dye-filled microcavity is an ideal platform to investigate rich nonequilibrium phenomena and the intricate relationship between thermalization and driven-dissipative processes. From Bose-Einstein condensation in near equilibrium systems \cite{Klaers2010b, Marelic2015,Klaers2010} to formation jitter during nonequilibrium quenched dynamics \cite{Walker2020}, these systems have provided important insights in the study of statistical  and dynamical properties of complex physical systems.

This primary focus of our work was to investigate the transport of light under the influence of different dynamical processes inside a dye-filled microcavity. In particular, the aim was to classify the transport in terms of the thermalization, driving, and disorder in the system. For an ordered but biased cavity potential, it is observed that there exist two distinct regimes of transport, viz. conductive and localized, depending on the thermalization and the driving rate in the system.
In the conductive regime, we observe that the light primarily propagates across the lattice through a series of dynamical phase transitions, whereby a cavity mode close to a lattice site undergoes condensation and becomes macroscopically occupied. The speed with which the light moves inside the cavity is qualitatively equal to the time taken for the mode at each lattice site to condense.
{In contrast, for poor thermalization or below-threshold driving of the molecules, the light in the cavity is locked in the localized regime, where no propagation of light is observed.}

Moreover, the effect of 
disorder in the {quasi 1D lattice potential} also plays a significant role in the transport of light inside a cavity. In both the limiting cases, where the disorder is either negligible or strong, nonequilibrium processes tend to become unimportant, and the light is in the conductive and localized regime, respectively, regardless of the thermalization. 
However, for weak disorder in the system, the propagation of light inside the cavity is stronger in systems that are better thermalized. {An important question here is whether a similar effect can be observed in two-dimensional  lattices in transverse plane of the cavity. While the thermalization properties are expected to be qualitatively similar, the spatial distribution of cavity modes under comparable disorder may not lead to significant localization of light. 
Additionally, beyond the nonequilibrium model considered here, a mean-field, nonlinear optics based approach \cite{Leeuw2013, Nyman2014,Wurff2014,Strinati2014,Alaeian2017} can be taken to study the dynamics of the lowest cavity modes.
This may allow one to investigate the role of effective photon-photon interactions \cite{Strinati2014,Alaeian2017} in the transport of light. Such interactions are often introduced upon changing the molecular reservoir by using Kerr nonlinearity and thermo-optical effects.}


A notable aspect of our study on transport and localization of light is that the different thermalization conditions and parameter regimes can be closely controlled, especially 
in physical implementation of the system based on photon gases inside multimode, dye-filled microcavities.
%
The structure and dimension of either the biased or disordered, random lattice potentials, as sketched in Fig.~\ref{fig_modes}(b), can be fabricated with minute precision on the transverse space of the cavity mirrors \cite{Walker2020b}. Further, the thermalization conditions can be controlled by tuning the cutoff frequency of the microcavity for a specific photon loss rate.
Therefore, the system can be readily prepared to allow for controlled propagation of light across several spatially distant optical modes or readily switch to a more localized light containing only a few modes.
An immediate result of our transport study is the possibility of forming a Bose-Einstein condensate at a distant location in the lattice potential. The lowest-energy mode can be made to condense via successive phase transitions in the lattice away from the initially pumped region. 
This was recently shown for simple lattices containing single and double well potentials \cite{Vlaho2019}. 
However, from the perspective of disordered potentials, the formation of condensates at a given location is a trade-off between the randomness and the dynamical processes. 
As such the nonequilibrium transport model could also be important in the study of topological transitions in random networks that behave as a Bose gas \cite{Bianconi2001,Bianconi2001b}, and could prove useful in studying evolution of complex systems.

\acknowledgements
We acknowledge financial support from EPSRC (UK) through the grants EP/S000755/1, the Centre for Doctoral Training in Controlled Quantum Dynamics EP/L016524/1 and the European Commission via the PhoQuS project (H2020-FETFLAG-2018-03) number 820392. We also thank Kiran E. Khosla and Hyukjoon Kwon for insightful discussions on the transport of light.


\begin{thebibliography}{99}
\bibitem{comment1} This definition is independent of the exact fraction of light intensity used to delineate the wavefront.
Any high percentage (90-99.9\%) of light is sufficient to define the intensity wavefront for further analysis.
\end{thebibliography}
\end{document}